# On the Internet, Nobody Knows You're a Dog… Unless You're Another Dog

**Ilyena Hirskyj-Douglas**
Aalto University
Espoo, Finland
ilyena.hirskyj-douglas@aalto.fi

**Andrés Lucero**
Aalto University
Espoo, Finland
lucero@acm.org

## ABSTRACT

How humans use computers has evolved from human–machine interfaces to human–human computer mediated communication. Whilst the field of animal–computer interaction has roots in HCI, technology developed in this area currently only supports animal– computer communication. This design fiction paper presents animal–animal connected interfaces, using dogs as an instance. Through a co-design workshop, we created six proposals. The designs focused on what a dog internet could look like and how interactions might be presented. Analysis of the narratives and conceived designs indicated that participants' concerns focused around asymmetries within the interaction. This resulted in the use of objects seen as familiar to dogs. This was conjoined with interest in how to initiate and end interactions, which was often achieved through notification systems. This paper builds upon HCI methods for unconventional users, and applies a design fiction approach to uncover key questions towards the creation of animal-to-animal interfaces.

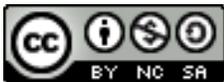

## INTRODUCTION

In the late 1970s personal computers got introduced into our homes: however, these early computers, originally only allowed humans to 'talk' to the computer. In the early 1990s with the introduction of the World Wide Web (WWW) people at home could communicate with each other evolving further complex usages such as virtual connected environments [32], wearable technologies [51] and advanced social media [7].

Along with humans, non-human animals (hereafter animals) have also been using computer technology for some time [15]. Taking a similar route to humans in Human–Computer Interaction (HCI), studies have increasingly investigated how animal(s) interact with technology, the design of these systems and the impact of these interactions [29]. This field has been coined Animal Computer Interaction (ACI) [29]. This recent growth in animal-computer research aligns with an increase in consumer products for animals [18]. This technology spans from animals that we keep in zoos and sanctuaries [6], towards animals that we domesticate as pets in our homes [14, 33, 49] and animals in the wild [23]. ACI research has seen technology systems designed for animals to both assist humans [5, 52] and to entertain animals [43] aimed towards increasing the animal-human bond [6, 49]. As there is a large number of animals living in domestic or caged situations, ACI has also investigated connecting owners to their dogs [41] and zoo visitors to the captive animals [6, 48]. With the annual ACI conference entering into its fifth year and animal-computing taking further root in HCI, the way in which animals use computers is evolving [15].

However, animal-computing unlike humans has yet to take the step of interconnectivity between users moving from the typical use in the 1970s of human-to-computer towards human-to-human communication via computers. Whilst researchers in ACI and ethology have pondered upon an interspecies internet allowing animals to talk to each other [42] as we progress through the technology age, the step that HCI has made of human-to-human communication via computers, has still yet to be investigated with animals.

This paper pushes forward both HCI and ACI by investigating what would these animal-to-animal interactions via the internet look like. In this paper we take dogs as a case example. As a 2018 PDSA PAW report[1] (a study of animals in the UK) suggests, out of 8.9 million dogs in the UK, 2.1 million were left alone in their owner's home for quite considerable lengths of time. This is not always progressive towards a dog's social welfare, which without being able to socialize may be inclined towards more negative and problem behaviors [46]. The consequence of poor socializing has been reported to be the second biggest issue concerning a dog's behavior due to the lack of dog-to-dog contact. Coupling this demographic with the increase of consumer systems for dogs, it is clear that technologies in the home that allow dogs to connect is a space open for investigation. This trend has been noticed within ACI with technologies to connect the owner to the dog via video calls both in research [13] and consumer products. However, primarily pets are left at home as the owner(s) are busy elsewhere such as at work. Thus, a dog-to-dog internet could potentially allow this interconnectivity when owners are not available, reducing the socializing issues and thus benefiting both the owner and more importantly the dog.

The following body of work uses the established HCI technique of speculative design/design fiction scenarios [9]. This method has also been employed previously in animal-computing with horses [34] and dogs [24]. As such, design fiction was chosen due to its previous use and its favorability to ideate around futuristic scenarios. These designs were developed and iterated over an open workshop on 'Designing Technology for Dog-to-Dog Interaction' at the Student Interaction Design Research (SIDeR)[2] conference. This workshop was informed through the first author's previous work on dog-technology for dogs within the home [14, 15, 16, 17, 18]. Here six designs are illustrated; *Companion-Dog*, *Virtual Walk*, *DogFlix*, *Tangiball*, *RopePull* and *Laser Collar*. This narration presents the first venture into what an animal-to-animal internet would be shaped like. Given dogs' socializing needs and that they have already been using technology systems for some time, three research questions are raised:

**RQ1**: What would a dog internet look like?

**RQ2**: What would these dog-to-dog interactions look like?

**RQ3**: How effective is the method of design research for animal- to-animal internet interactions?

These questions are considered in a design fiction study focusing on designing technologies for dog-to-dog interactions. This study is especially pertinent given the relatively new interest in animals' connectivity and entertainment style interfaces for dogs (and other animals) [14]. This research forefronts a change in ACI to move from requiring an animal owner for interaction into an animal-led system [18].

## RELATED WORK

ACI investigates how animals interact with technology, among other things, and the design of these technologies [29]. Whilst initially ACI took primary influences from HCI contexts [17, 39] centering around the usability and user experience of the technology [49] towards interactive devices, [25, 45] frameworks have since been developed across areas such as interaction design [45, 30], game design [12, 37, 49, 50] and ubiquitous computing [29]. Recently, as the motivation of animal-computer technologies has often been welfare based [29], ACI also draws from animal behavior and cognition [12, 18, 20, 35, 44] and animal welfare [22, 42]. Part of this conversation includes investigating the role and position that the technology plays within the interaction space towards human–animal interactive systems [21, 30, 49] and animal-computer systems [14]. ACI also functions to develop playful systems for animal entertainment [18, 49] and wellbeing [12, 45] where consumer products predominantly exist in the pet sector.

### Dog-Computer Interaction

A sub field of ACI when focusing upon dogs is Dog-Computer Interaction (DCI). Dogs are unique in DCI as they can take on varying roles from both a working dog [52] and as a pet [18, 49]. Whilst dogs can occur in both positions, there is a difference in the human's motivation towards these technologies in regards to the requirements and animal-centeredness; whose needs the technology is meeting (human or dog) and the requirement to be dog centeric [14]. Drawing from this, the main concern in ACI is understanding the animal to be able to design tools and methods that aid in researching animal interactions with technology [3, 30]. As animals cannot communicate effectively using vocal language, the default means of interpretation in ACI is instead to use behavioral and physiological measurements [12, 25] often deduced through observations [36].

---

[1] PDSA Animal Wellbeing Report https://www.pdsa.org.uk/media/4371/paw-2018-full-web-ready.pdf

[2] SIDeR. http://sider18.aalto.fi/

In DCI interfaces, a dogs response can be derived from facial reactions such as eye movements [44], nose movements [20, 52] and head movements [14, 17, 47]; behaviors such as biting [19], pulling [40], pushing buttons [12, 39], touch screens [52] and posture analysis [31]; and biological responses such as heart and respiration rates [31] and hormone levels [12]. These behaviors and biological responses have been used within DCI systems to allow the dog to feed-back to a system such as operating nose plate interfaces [20], bitable pulleys and buttons [19, 40], paw activated buttons [12, 39], proximity [18], and haptic vests [19, 26, 5].

**Designing for Dog-Driven Interfaces**
Nonetheless, so far DCI interactions have all been driven to connect the dog to a computer interface for human data collection. These include alerting the human as a call for help [40, 52], monitoring the dog [41], collecting information about a dog's physiology [26, 44], training the dog [33] or using the dogs enhanced senses such as smell [20, 30]. Parallel to this, there has been games and technologies for dogs, but these are human-initiated interactions [49]. It is only recently that researchers have investigated dog-led interactions [18] paving the way for dogs to use and control computer technologies as they desire. Nonetheless, this leaves a gap in research to investigate dog-led technologies. Part of this conversation over the dog's interaction method includes to what extent a dog can be involved within the design process [24, 15].

While Mancini & Lehtonen [30], Hirskyj-Douglas et al. [16], Westerlaken & Gualeni [49] and, Lawson et al. [24] all advocate for participatory design with dogs, these are taken through different approaches. Mancini & Lehtonen [30] recently advocated for widening what we think of participation based on indexical semiosis (i.e., establishing contextual associations between consistently co-occurring events), volition (i.e., the cognitive process by which an individual decides on and commits to a particular course of action), and choice. Hirskyj-Douglas et al. [16], whilst similarly advocating for choice, argue that for a dog to be a participant, the dog's interaction should be dog-centric where the dog's decisions directly impact the end-design. This view, unlike Mancini & Lehtonen [30] who see the dog as inherently 'taking part' in the study as an act of participation, instead see participatory research as currently conceptual due to it being unknown if the dog understands, or could understand, the interaction taking place. Westerlaken & Gualeni [49] take a different approach, drawing from sociology works of Bruno Latour and Donna Haraway, to argue towards interpretation from the human perspective as a form of participation itself, much like Mancini & Lehtonen [30]. Alternatively, drawing closer to Hirskyj-Douglas et al.'s [16] and Mancini & Lehtonen's [30] argument for participation, Lawson et al. [24] believe for dogs to be considered a participant, the system should reflect and represent the values of the stakeholders. This is conducted in a shared manner between human and dog. Still, how this is implemented in dogs has not yet been clarified.

All these models on participation however share the same challenge of including the dog end-user as a stakeholder just varying on levels of representation, interpretation and contribution. Thus, this is a procedure of mitigation as allowing the animal to decide upon the properties and affordances in the process can be ambiguous and unregulated. This is due to it being unknown within these shared experiences as what constitutes a positive and playful experience beyond our acknowledgement [16, 21]. As such, there are clear challenges within DCI towards both how the dog interacts and has productive power within the design process through participation.

**DESIGN FICTION IN ANIMAL-COMPUTING**
One method that has been used in ACI for designing with horses [34] and dogs [24] is design fiction. Design fiction presents "fantasy prototypes" of what is plausible in the near future [4]. These prototypes are often provocative: scenarios narrated through designed artifacts as a way to facilitate and foster debates. Often seen as a social object for communication, these designs are hoped to provide substance enough to enable meaningful conversations [43].

However, what is defined as a meaningful conversation, or to whom, is uncertain especially in animals as Hirskyj-Douglas et al. [15] and the ACI Manifesto note [29]. Design fictions can involve many practices, such as short stories, objects, prototypes, narratives and films [4]. Lindley and Coulton [27] have proposed design fiction to fit into three delineations: 1) creating a story world, 2) prototyping within that story world, and 3) creating a discursive space, where 'something' may mean 'anything'. Useful to ACI, design fiction as a method helps us to situate a new technology within a narrative, that is dog internet, to help us grapple with the surrounding questions of ethics, values, social perspectives, causality, politics, psychology and emotions [27]. This issue of ethics and inclusion has been tackled in ACI using design fiction as a way of including the animal as a participant by seeking to design with their goals in mind. This aims, if only speculatively, to provide meaningful interactions [24, 34].

Lawson et al. [24] used design fiction with dogs to see what a dog internet would look like and what it would be used for (Figure 1). Lawson et al.'s [24] design fictions highlight the need to develop devices for animals whilst acknowledging the tension between measuring an animal need and not knowing what these are when the user is not human. Lawson et al. [24] work ends on suggesting more speculative design to aid understanding whilst appreciating the lack of knowledge.

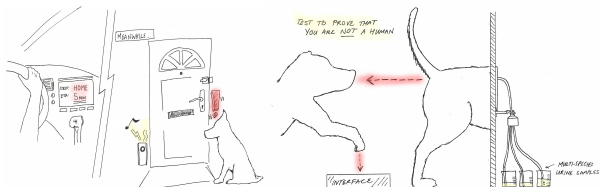

**Figure 1: Lawson et al. [24] design fictions for a dog internet. Left image shows a dog being notified of its owner arrival home, right image shows a dog using a dog butt interface as a smell CAPTCHA test.**

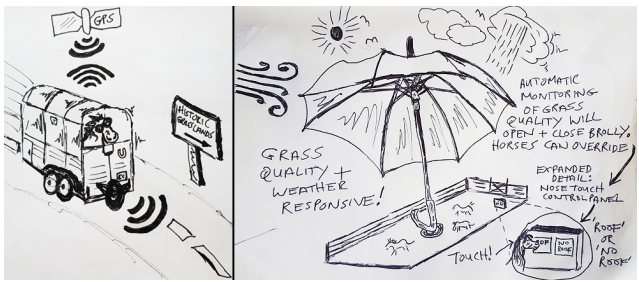

**Figure 2. North's [34] design fiction sketches: Left image . showing one horse controlled carriage for transport and the right image demonstrating an umbrella roof.**

In a similar stance North [34], who investigates design fictions for horses (Figure 2), found design fictions as a useful tool in ACI to explore the 'otherness'. North's [34] design fictions, primarily looked at how technology can support a horse's daily life such as an umbrella to shield the horse from the sun and a form of transport in the shape of an automated horse trailer to bring the horse back home (Figure 2). Part of this required designers to familiarize themselves with ethology to provide both the initial contribution and for the evaluation. North's [34] designs were created when he co-designed with his horses, although it is not explicit how the horse was a participant. Lawson et al. [24] did not co-design with animals, but included multiple designers within the fictions' creation. As such, design fiction has been found useful in ACI for proposing systems that allow animals to connect to technology. However, whilst Lawson et al. [24] have investigated what a dog internet would look like, this early computing only connects the dog to machines and not to other dogs. This leaves the question of what an animal-to-animal internet would look like underexplored.

## WORKSHOP METHOD

We organized an open workshop on 'Designing Technology for Dog-to-Dog Interaction' during the Student Interaction Design Research (SIDeR) conference in May 2018. There were twelve participants in total; nine human attendees, two human facilitators and one dog. The dog was involved in the co-design process through playful interactions, as a stakeholder [24, 34], with the other human participants in line with ACI research methods [49, 50]. As these speculations involved designing for a user with different requirements, it was imperative that the participants have familiarity with their users (dogs) to have a level of understanding and weight of importance towards their decisions. We had a mix of human participants, most of who had previously lived with dogs (6/11) and other animals including pets and farm animals (9/11). Two participants had never owned an animal, but had a growing interest in ACI. Especially for these participants but generally for everyone, the dog was present to play with using the relevant toys to the activity. This allowed the dog to help the participants begin to think about the speculations from a more concrete starting point. Additionally, each group had at least one current/past dog owner, and a current/past owner of another species. The participants' background varied from product/ interaction design to general IT and engineering with all participants being intermediate-expert in technology.

Inspired in co-design, the participants were involved in ideating futuristic scenarios in the context of ACI using an adaptation of the *dialogue-labs* and World Café methods. The *dialogue-labs* [27] combine the use of *process*, *space* and *materials* in a structured way and provides a clear step-by-step procedure for a two-hour idea-generation session in which participants first work in pairs and then as a whole group [28]. The World Café consists of five main elements that can help host large group dialogue (i.e., *setting, welcome and introduction, small group rounds, questions,* and *harvest* [3]). This section describes how we applied these two methods, and what kinds of task and materials were used for sensitizing and facilitating ideation.

### Tasks

The overall task of the workshop was to think of ways to encourage dog-to-dog interaction. Three subtasks were defined in order to encourage the participants to approach the topic from different angles. The tasks draw attention to (1) Screen and Tracking Systems, (2) Haptic and Wearable Systems, and (3) Tangible and Physical Systems. These classifications of ACI systems were chosen based upon Hirkyj-Douglas et al.'s [15] and Jukan et al.'s [22] categorization of technologies in ACI. For each task the participants were asked to think about the following:

- How could dogs use the task subject to communicate with each other remotely?
- How would this interaction start and end?
- How could the dogs using the system have symmetries?
- What would these communications look like in terms of how a dog could interact ordinarily?

### Materials

Each task was accompanied with video material[4], pens, pencils, A3 paper, robotic dog toys[5], dog toys (ropes and balls), and sticky notes to support ideation. The purpose of these videos, some of which were about dogs, was to provide examples to extend participants' knowledge on ACI, and open new avenues for ideas. There were no rules regarding how each trio should work together. The instruction was simply to *ideate together*. Groups however were encouraged to play with the technology and the dog actively during their idea generating which each group did.

For Task 1, the video consisted of North et al. [35] speaking about their work on an automated horse tracker for behavioral detection. This was followed by a video explaining how dolphins use the acoustically-operated touchscreen device, Echolocation Visualization and Interface System (ELVIS) [2]. The concluding video for this task presented Rossi et al. [41] explaining how to use a program to video call a dog to do commands. These videos were chosen as they vary from traditional screen interaction (video call) to a senses-focused screen interaction (echo location) giving examples of tracking methods.

For Task 2, the video began with Alcaidinho et al. [1] explaining how haptic systems have been used to give dog commands for training purposes. This was followed by Paci et al. [36] talking on wearable biotelemetry devices for cats and other animals. Lastly, Jackson et al. [19] speaks about research being conducted in Georgia Tech with dogs using haptics and wearables. This video focuses upon wearable vests that allow the dog to 'talk' through pressing certain buttons. These videos aimed to demonstrate that haptics can be used for two-way communication to not only allow the dog to input into technology, but for the technology to react to this input and feedback. This included more traditional wearable technologies (such as GPS collars [36]) but also more dog-initiated technologies (bite-able buttons [19]).

For Task 3, the video consisted of Mancini [29] speaking about physical button, and rope pull alarms systems for assistant dogs at the Open University. This was followed by Westerlaken et al. [49] explaining their game where a dog has to climb physical objects to unlock the level and thus a treat. These videos were chosen as they showed the roles that physical and tangible systems can take in both work and play.

### Procedure

Each two-hour session started with a 15-minute sensitizing phase. To aid in designing dog-to-dog technologies, we first provided a brief history of ACI. In order to build common ground, we presented examples of commercially-available dog- and cat-computer interaction (e.g., FitBark [6], PetChatz[7], PetCube[8]) products. Once familiar with current dog-to-dog technologies, the hope was that participants would start thinking beyond the status quo.

After the sensitizing phase, the group of nine participants was divided into three trios. Each trio chose one of three locations with a task to ideate dog-to-dog speculative interactive designs, and document them. After 15 minutes, they changed to a different task. Three ideation rounds were performed. During each round, groups generated and documented their ideas on sticky notes and drawings on the paper. Instead of clearing the table at the end of each round, and in the spirit of World Café, groups were asked to summarize their ideas by leaving their sticky notes and ideas behind to aid incoming groups in making sense of the activity once they arrived to a new station. Groups were allowed to leave or take their other papers (main designs) with them to enable them to further iterate on their designs, or start afresh. This was done to enable the final designs to incorporate the different tasks.

---

[3] World Café. http://www.theworldcafe.com/key-concepts-resources/world-cafe-method/
[4] Dog-to-Dog Interfaces: Tangible & Physical https://youtu.be/8rU2FWT5Koc Haptic and Wearable https://youtu.be/NiG0owYOg9M Screens & Tracking https://youtu.be/-xAF9-GPMVw

[5] Sphero. https://www.sphero.com/
[6] FitBark. https://www.fitbark.com/
[7] PetChatz. https://petchatz.com/
[8] PetCube. https://petcube.com/

After a five-minute break, the pairs presented the ideas they thought as their best ones for 15 minutes. This was followed by the groups narrating their journey around the different stations, speaking about how their designs changed and their design decisions. After these group activities, the group as a whole spoke about design challenges that they faced, key problems they highlighted and discussed the general theme of the workshop topic as a whole.

**Analysis**

The co-design workshops resulted in a wide range of ideas of different levels of fidelity and focusing on different aspects of dog- to-dog interaction. We analyzed the data bottom up, first transcribing the group discussions and identifying individual ideas. To objectively measure the quality of ideas, we then filtered out the ideas which a) did not allow, or led to, two dogs communicating, and b) have a start and end point to the interaction. We then chose the best ones, which are illustrated below through the sketches that the participants made during the workshop. This is accompanied by text describing the design fictions taken from the participants narration, their sticky notes left on the table and keywords/writing around the drawings. The best sketches that capture the essence of the ideas created from a congregated perspective are presented here. The two facilitators grouped the ideas according to commonalities and enriched the illustrations with keywords. This led to the identification of six design fictions (i.e., *Companion-Dog*, *Virtual Walk*, *DogFlix*, *Tangiball*, *RopePull* and *Laser Collar*). The rest of the analysis was an iterative process of going between data, writing, and modeling.

**RESULTING DESIGNS**

**Design Fiction #1 – Companion-Dog**

With dogs often being left home alone, they can end up feeling sad and unaccompanied. *Companion-Dog* aims to provide a solution to this problem through a dog plush toy (Figure 3). When this toy is picked up, it alerts the other *Companion-Dog* toy in the dogs' friend's house through a series of 'woof' sounds. If the other dog would like to play, he/she also picks up the *Companion-Dog* toy. A connection/interaction is then established when both dogs drop their *Companion-Dog* toys in the basket. These dogs can now see a projection of each other and communicate this way. If the dogs would like to play further there is a button which can be pressed with a paw. Upon being pressed, a ball will then be shot towards the other dog allowing the dog to play 'fetch' with each other stimulating the role a human would normally take.

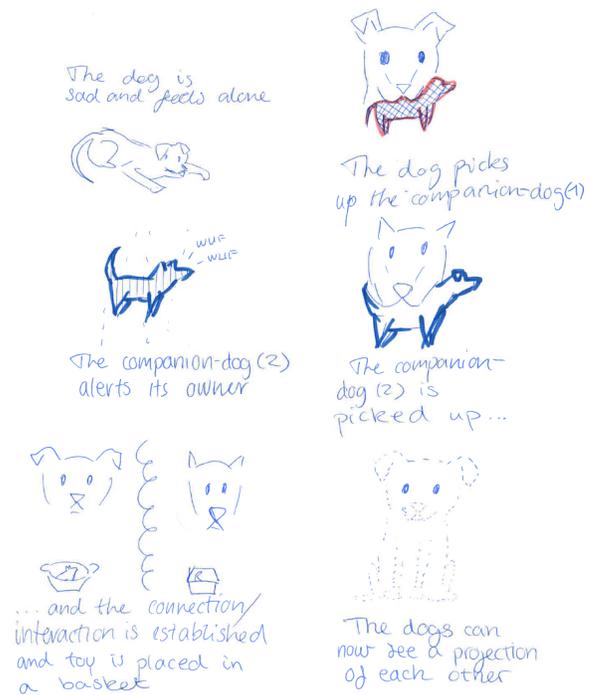

**Figure 3. Sketch of Design #1 Companion-Dog.**

**Design Fiction #2 – Virtual Walk**

When dogs go for walks outside, they often encounter other dogs to walk and play with.

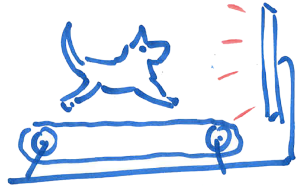

**Figure 4. Sketch of Design #2 Virtual Walk.**

Virtual Walk (Figure 4) allows the dog to 'walk' and 'run' with his/her friends whilst in the home. This device is made up on a treadmill which the dog can run/walk on and a screen to see their friends also running/walking at the same time.

**Design Fiction #3 – DogFlix**

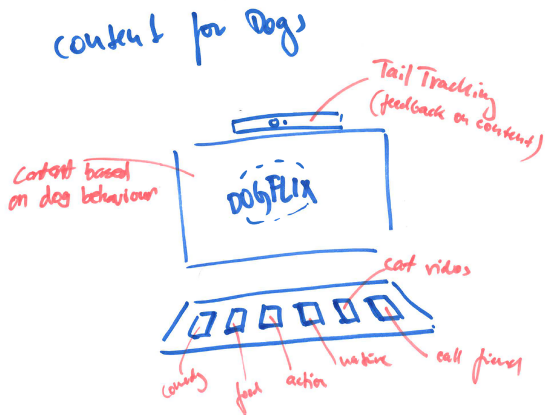

**Figure 5. Sketch of Design #3 DogFlix.**

Recent research has invested building robots to allow dogs to interact with screens [18] alongside a recent increase in TV channels made for dogs. These devices however are either for sole dog use or do not provide a way for dogs to watch screens together. DogFlix (Figure 4) provides content for dogs to allow joint viewing. DogFlix consists of a TV device, a tracking module and a dog keyboard. When the dog is present, DogFlix turns on (otherwise remaining blank) and shows content to the dog. This content is based upon the dog's behavior, where his/her feedback is collected through his/her tail wags (equivalent of a 'like'). The dog keyboard has several buttons which allow the manual changing of the content to comedy, food, action, cat videos and nature. This keyboard also has a button to allow the dog to call a friend, who will then also be shown on the TV screen and the content shared allowing joint viewing.

**Design Fiction #4 – Tangiball**

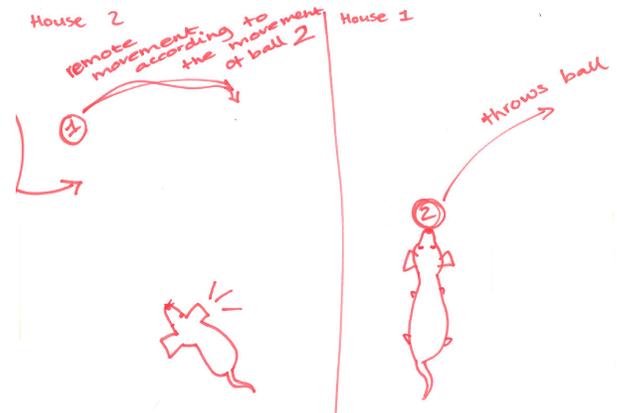

**Figure 6. Sketch of Design#4 Tangiball.**

Similar to Design#1 Companion-Dog, this device allows for two dogs in different homes to play fetch with each other. With *Tangiball* however, this product allows two dogs to play with balls that remotely move the same whilst seeing holograms of each other. For example, this allows one dog in their home to pick up *Tangiball* and throw it, throwing the *Tangiball* in the second dogs home. As a result, this allows two dogs to play with each other simulating two dogs playing with a ball at the park.

**Design Fiction #5 – RopePull**

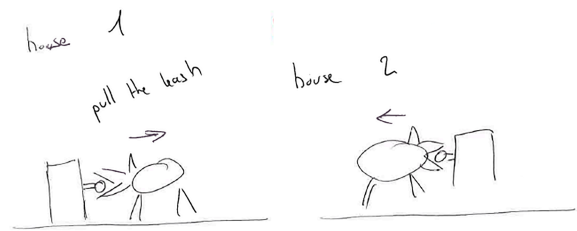

**Figure 7. Sketch of Design #5 RopePull.**

Dogs are often seen playing 'tug-of-war' games with ropes. These games are described in Resner's [39] work as being an ideal symmetric interface and thus used by Robinson et al.'s [40] research as an interface mechanism. In *RopePull*, like *Compainon-Dog* and *DogFlix,* when the first dog interacts with the technology a notification will be sent to the second dog to invite them to play. This notification is 'sent' by the dog when he/she pulls on the rope attached to the box. This will then wiggle the rope in the second dogs home alerting the dog that his/her friend wants to play. Once both dogs have 'bitten down' on the rope the tug of war game begins. The strength of the 'tug' from each dog is sent through the rope to allow for long distance interaction with the tangible rope object.

**Design Fiction #6 – Two Dogs Meet at a Bar: Laser Collar**

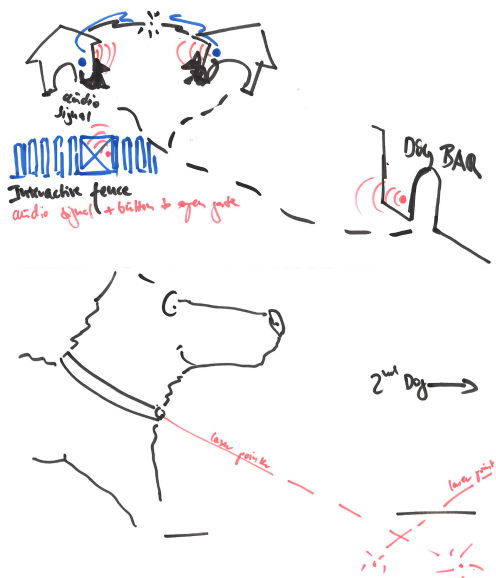

**Figure 8. Sketch of Design #6 Laser Collar**

With the rise of GPS systems in phones and cars, finding directions to meet each other has become a regularly used system. *Laser Collar* (Figure 6) aims to solve this issue with dogs by providing a way for dogs to arrange to meet each other, in this scenario meeting at the 'dog bar'. The *Laser Collar* system has two parts; the collar itself, and a fence system to provide further security. The system is started by one dog calling the other dog so that both dogs are notified of the other's intention (symmetric interaction). The laser, which is attached at the front of the dog's collar, is then activated which points towards the 'dog bar' guiding the dog. As the dog gets closer to the 'dog bar' they will get further signals via audio sounds from the fence. As they reach the 'dog bar' premises, she/he must pass this 'interactive fence'. This fence detects if the dog is friendly, and if so, lets the dog(s) in the premises where they can follow their collar to the 'dog bar' entrance and meet their friend(s).

## DISCUSSION

Over the previous pages, we have used the method of design fiction to describe six dog-to-dog speculative interactive systems; *Companion-Dog, Virtual Walk, DogFlix, Tangiball* and *Laser Collar*. However, these designs were used, as Lindley and Coulton [27] noted, to stimulate a discursive space. Certain key themes were noted when the creators of the sketches reflected upon their design decisions and explained their design fictions to each other. The findings from this study are relevant in two domains: the first is in the designs themselves towards creation of technology for dogs to interact with other dogs through the internet, the second is through the interaction mechanisms that the designs reflect out.

### Technology Creation for Dog-to-Dog Interaction

Drawing from the designs, it is evident that they all involved repurposing key elements of an interactive object which dogs use in everyday life such as lasers in *Laser Collar*, tug toys in *RopePull*, balls in *Tangiball* and dog toys in *Companion-Dog*. Current work in ACI and ethology has begun to look at what these devices mean for dogs [12, 17, 30] stating that the interfaces used should reveal its functionality through its affordances, where the artifacts need to be self-discoverable and intuitive [39]. This requisite for constant affordances is self-evident in current dog toys where the use is implicit in the design and as such have been repurposed to be used in technology interfaces [11, 40]. However, ACI has not yet investigated how objects should be repurposed, and the impact thereof.

To map device affordances systematically, we need methods to measure and infer responses from animals. Methods such as eye gaze and posture [44], biological responses [12], and carer/owner feedback [14] have previously been used. Within our design fiction scenarios, this response mechanism was either spoken about within an automated behavioral recognition such as 'tail wags' for 'liking' in *DogFlix*, or through the dog choosing to partake within the activity such as 'dropping the toy' with *Companion-Dog*. This choice to 'take part' has been spoken about within ACI as a form of 'consent' [29, 30] through giving the dog more autonomy [18]. This was thought about within the design fictions through the dogs moving novel objects (dropping the toy/ not using the device to end the interaction) or walking away (from the screen).

Looking towards how HCI designs with unordinary users, such as children, the key here could be to investigate into new novel feedback that assists the user. For instance, one method investigating in children in Child-Computer Interaction (CCI) is the 'Fun Scale' which was used to measure fun of interactive products rather than the traditional survey method used with adults [38]. In this way, it is clear that this adaptation of traditional HCI methods also needs to be done in ACI. These design fictions highlight the need for more methods to be dog-initiated.

### Interaction Paradigm for Dog-to-Dog Interaction

Often discussed when talking over these designs, was that the connection for dog-to-dog interactions was in two locations with the technology as the mediator. As such, there was a need on both sides to interpret the interaction. In the words of a participant taking part within the workshop "*What can you interpret from tracking the dog?*". In ACI, questions have been raised reflecting upon Norman's gulfs of interactions upon what it means for a dog to understand technology [10, 15]. Within these design ideations, participants raised this question as a philosophical problem of how you allow dogs to play with other dogs knowing that they are playing with each (asynchronous interactions) mediating towards a dog's intentions. As mentioned by North [34] all animals (here he refers to humans and non-humans) are interacting somewhere along a continuum of awareness. Here he stresses that unless an individual is able to contextualize their interaction, there will be a lack of 'system awareness' and less explicit interactions. It was this awareness and intention that the participants in our workshop wanted to explicitly gather from the dogs using their systems.

Whilst Friel et al. [10] believe that a dog's intention (*the gulf* to use Norman's terms) can be measured using terminologies. Alternatively, Hirskyj-Douglas et al. [15] believe that this loop is not closed where the bottom half of the loop (the *gulf of evaluation*) is currently unknown as whilst an animal's interaction can be captured in some sense, what meaning this interaction has to the animal (their interpretation of the interaction) has yet to be determined [15]. In this case, what a dog would expect to get from another dog in a computer system and its intentions from using such system is unknown. Especially as the meaning and experience is interpreted through human acknowledgement [21]. Thus, ACI is not primarily focused on designing complete systems with cognitive reasonings but instead examining components within these gulfs through acknowledging the unknown spaces [15]. This same paradigm also transpires in HCI systems, e.g., young children, where it can be difficult to determine the user interaction(s) [38].

**Further Work**

As narrated here through both the literature and the participants' queries, there are clear gaps in research to explore an animal's interactions, and what it would mean for a dog to be connected to another dog. This delineation raises the question of what would dog communication look like. Whilst within our design fictions these are mostly visual or auditory, as noted by Lawson et al. [24] these can also be olfactory where dogs communicate biological and emotive intentions [20]. Within the design fictions, participants often spoke about having numerous methods of the dog interacting to allow this choice of mechanism highlighting that there is more here to be investigated. One of the key confines of this study was both the limited familiarity that the co-designers had with the dog subjects, the few designers themselves and the limited number of dogs. Whilst these were restrictions within the environment, without this initial study there would not be data to support more experimental techniques towards ideating what a dog-to-dog system would look like. As such our future work will iterate over again with experts across various fields (such as animal behavior and CCI) and multiple dogs to provide a broader perspective on the issues raised.

**Implications for Human–Computer Interaction**

While this work is based in the field of ACI, it also has implications for HCI. Firstly, addressing specific user populations, such as those with special abilities and affordances including non-verbal communication and limited cognition, has been a large area of interest for HCI researchers. Thus, while in ACI we design for and talk about systems for animals, the findings and outcomes of such discussion progress current methods and systems for specific user groups forming a symbiotic relationship between ACI and HCI. Secondly, when using an HCI method with animals, the fundamental assumptions that are made with humans are highlighted coercing the designer to make conscious decisions about the held assumptions. This includes what the user values, how we draw these evaluations, and the best way to meet the users need and requirements. Here we examine the value of design fiction to reveal and draw out the assumptions within the design process, questioning what an interaction encapsulates for a user and how as a designer we can design for the 'other'.

**CONCLUSION**

As the explosion of the internet and a multitude of devices have allowed for humans to be constantly connected, we speculate that the evolution from human–machine to human–human mediated via machine will happen with non-human animals. This ideation of animal–animal interactions over the web is not a new occurrence, but has yet to be implemented beyond animal–machine interfaces. In this paper we use dogs to explore what dog-to-dog technology mediated interactions would look like. This choice is taken as dogs are currently the key player within animal- computing, taking roles within our homes and workplaces. Furthermore, dogs are repeatedly left alone causing potential behavioral issues from lack of socializing creating a need for in-home technology. This study conducted through nine participants in a workshop, uses a design fiction method to produce six scenarios and then further reflecting upon this discursive space. Through this co-design, participations raise questions on the technology used to mediate the interaction. These focus on both a human's ability to measure an animal's behavior, and an animal to feed back. Drawing from this, further queries are noted around what it means for a dog to interact, if this interaction needs to be asynchronous and what would this interaction look like. We suggest that this problem be tackled through drawing upon more well-researched fields in HCI such as Child-Computer Interaction (CCI). In conclusion, six design fictions for dog-to-dog interactions are presented; *Companion-Dog*, *Virtual Walk*, *DogFlix*, *Tangiball*, *RopePull* and *Laser Collar*. In the field of HCI, much is made of design fiction to help stimulate conversations around future technology. Whilst the dialogues around dog-to-dog interactions hold note towards animal-computing systems, as demonstrated, the conversation held these points of interest (such as feedback loop, agency and technology mediators) ring true towards current HCI matters. It is in this way that we encourage designers working in this area to draw and engage in currently deployed HCI. Overall, this study contributes new understandings in the co-design of computer interactions for animal users to connect to other animals, highlighting key questions for both those in ACI and HCI.

## ACKNOWLEDGMENTS

We would like to thank the human and dog participants who took part in the workshop aiding in growing this idea further and Soledad Paz and Kashyap Todi for editing and proofreading this document.

## REFERENCES

[1] Joelle Alcaidinho, Giancarlo Valentin, Gregory D. Abowd, and Melody Jackson. 2016. Training collar-sensed gestures for canine communication. In Proceedings of the Third International Conference on Animal-Computer Interaction (ACI '16). ACM, New York, NY, USA, Article 17, 4 pages. DOI: https://doi.org/10.1145/2995257.3012020

[2] Mats Amundin, Josefin Starkhammar, Mikael Evander, Monica Almqvist, Kjell Lindström and Hans W. Persson. 2008. An echolocation visualization and interface system for dolphin research. The Journal of the Acoustical Society of America. 123, 2. 1188-1194. DOI: https://doi.org/10.1121/1.2828213

[3] Fredrik Aspling and Oskar Juhlin. 2017. Theorizing animal-computer interaction as machinations. Int. J. Hum.-Comput. Stud. 98, C (February 2017), 135-149. DOI: https://doi.org/10.1016/j.ijhcs.2016.05.005

[4] Mark Blythe. 2014. Research through design fiction: narrative in real and imaginary abstracts. In Proceedings of the SIGCHI Conference on Human Factors in Computing Systems (CHI '14). ACM, New York, NY, USA, 703-712. DOI: https://doi.org/10.1145/2556288.2557098

[5] Ceara Byrne, Larry Freil, Thad Starner, and Melody Moore Jackson. 2017. A Method to Evaluate Haptic Interfaces for Working Dogs. Int. J. Hum. Comput. Stud. 2017, 98, 196– 207. DOI: https://doi.org/10.1016/j.ijhcs.2016.04.004

[6] Marcus Carter, Sarah Webber, and Sally Sherwen. 2015. Naturalism and ACI: augmenting zoo enclosures with digital technology. In Proceedings of the 12th International Conference on Advances in Computer Entertainment Technology (ACE '15). ACM, New York, NY, USA, Article 61, 5 pages. DOI: https://doi.org/10.1145/2832932.2837011

[7] Luiz Henrique C.B. Cavalcanti, Alita Pinto, Jed R. Brubaker, and Lynn S. Dombrowski. 2017. Media, Meaning, and Context Loss in Ephemeral Communication Platforms: A Qualitative Investigation on Snapchat. In Proceedings of the 2017 ACM Conference on Computer Supported Cooperative Work and Social Computing (CSCW '17). ACM, New York, NY, USA, 1934-1945. DOI: https://doi.org/10.1145/2998181.2998266

[8] Nick Couldry. 2012. Media, Society, World: Social Theory and Digital Media Practice. London: Polity Press. p. 2. ISBN 9780745639208.

[9] Anthony Dunne and Fiona Raby. 2013. Speculative everything: design, fiction, and social dreaming. MIT press.

[10] Larry Freil, Ceara Byrne, Giancarlo Valentin, Clint Zeagler, David Roberts, Thad Starner, and Melody Jackson. 2017. Canine-Centered Computing. Foundations and Trends® in Human–Computer Interaction. 10, 2. 87-164.

[11] Fiona French, Clara Mancini, and Helen Sharp. 2015. Designing Interactive Toys for Elephants. In Proceedings of the 2015 Annual Symposium on Computer-Human Interaction in Play (CHI PLAY '15). ACM, New York, NY, USA, 523-528. DOI: https://doi.org/10.1145/2793107.2810327

[12] Annika Geurtsen, Maarten .H. Lamers and Marcel .J.M. Schaaf. 2015. Interactive Digital Gameplay Can Lower Stress Hormone Levels in Home Alone Dogs: A Case for Animal Welfare Informatics. In 14th International Conference on Entertainment Computing (ICEC 2015). 238 -251. DOI: https://doi.org/10.1007/978-3-319-24589-8_18

[13] Jennifer Golbeck and Carman Neustaedter. 2012. Pet video chat: monitoring and interacting with dogs over distance. In CHI '12 Extended Abstracts on Human Factors in Computing Systems (CHI EA '12). ACM, New York, NY, USA, 211-220. DOI: https://doi.org/10.1145/2212776.2212799

[14] Ilyena Hirskyj-Douglas, Janet C. Read, and Brendan Cassidy. 2017. A dog centred approach to the analysis of dogs' interactions with media on TV screens. Int. J. Hum.-Comput. Stud. 98, C (February 2017), 208-220. DOI: https://doi.org/10.1016/j.ijhcs.2016.05.007.

[15] Ilyena Hirskyj-Douglas, Patricia Pons, Janet C. Read and Javier Jaen. 2018. Seven Years after the Manifesto: Literature Review and Research Directions for Technologies in Animal Computer Interaction. Multimodal Technologies and Interaction.2, 2. DOI: https://doi.org/10.3390/mti2020030

[16] Ilyena Hirskyj-Douglas, Janet C. Read, Brendan Cassidy. 2015. Doggy Ladder of Participation. British Human Computer Interaction (BHCI). Animal Computer Interaction.

[17] Ilyena Hirskyj-Douglas, Hulan Luo and Janet C. Read. 2014. Is My Dog Watching TV? In Proceedings of the NordiCHI'14—Workshop on Animal-Computer Interaction (ACI): Pushing Boundaries beyond "Human", Helsinki, Finland.

[18] Ilyena Hirskyj-Douglas and Janet C. Read. 2018. DoggyVision: Examining how Dogs (Canis Lupus Familiaris) Interact with Media Using a Dog Driven Proximity Tracker Device. Anim. Behav. Cognit. 2018. 5, 4. DOI: https://doi.org/10.26451/abc.05.04.06.2018

[19] Melody Moore Jackson, Clint Zeagler, Giancarlo Valentin, Alex Martin, Vincent Martin, Adil Delawalla, Wendy Blount, Sarah Eiring, Ryan Hollis, Yash Kshirsagar, and Thad Starner. 2013. FIDO - facilitating interactions for dogs with occupations: wearable dog-activated interfaces. In Proceedings of the 2013 International Symposium on Wearable Computers (ISWC '13). ACM, New York, NY, USA, 81-88. DOI: https://doi.org/10.1145/2493988.2494334

[20] Olivia Johnston-Wilder, Clara Mancini, Brendan Aengenheister, Joe Mills, Rob Harris, and Claire Guest. 2015. Sensing the shape of canine responses to cancer. In Proceedings of the 12th International

[20] Conference on Advances in Computer Entertainment Technology (ACE '15). ACM, New York, NY, USA, Article 63, 4 pages. DOI: https://doi.org/10.1145/2832932.2837017

[21] Ida Kathrine Hammeleff Jørgensen and Hanna Wirman. 2016. Multispecies methods, technologies for play. Digital Creativity, 27, 1. 37-51. Vancouver. DOI: https://doi.org/10.1080/14626268.2016.1144617

[22] Admela Jukan, Xavi Masip-Bruin, and Nina Amla. 2017. Smart Computing and Sensing Technologies for Animal Welfare: A Systematic Review. ACM Comput. Surv. 50, 1, Article 10 (April 2017), 27 pages. DOI: https://doi.org/10.1145/3041960

[23] Hiroki Kobayashi, Kana Muramatsu, Junya Okuno, Kazuhiko Nakamura, Akio Fujiwara, and Kaoru Saito. 2015. Playful rocksalt system: animal-computer interaction design in wild environments. In Proceedings of the 12th International Conference on Advances in Computer Entertainment Technology (ACE '15). ACM, New York, NY, USA, Article 62, 4 pages. DOI: https://doi.org/10.1145/2832932.2837012

[24] Shaun Lawson, Ben Kirman, and Conor Linehan. 2016. Power, participation, and the dog internet. Interactions 23, 4 (June 2016), 37-41. DOI: https://doi.org/10.1145/2942442

[25] Ping Lee, David Cheok, Soon James, Lyn Debra, Wen Jie, Wang Chuang, and Farzam Farbiz. 2006. A mobile pet wearable computer and mixed reality system for human–poultry interaction through the internet. Personal Ubiquitous Comput. 10, 5 (July 2006), 301-317. DOI: http://dx.doi.org/10.1007/s00779-005-0051-6

[26] Germain Lemasson, Dominique Duhaut, and Sylvie Pesty. 2015. Dog: Can You Feel It? HCI'15. England.

[27] Joseph Lindley and Paul Coulton. 2015. Back to the future: 10 years of design fiction. In Proceedings of the 2015 British HCI Conference (British HCI '15). ACM, New York, NY, USA, 210-211. DOI: http://dx.doi.org/10.1145/2783446.2783592

[28] Andrés Lucero, Kirsikka Vaajakallio, and Peter Dalsgaard. 2012. The dialogue-labs method: process, space and materials as structuring elements to spark dialogue in co- design events. CoDesign 8, 1 (2012), 1–23. DOI: http://dx.doi.org/10.1080/15710882.2011.609888

[29] Clara Mancini. 2013. Animal-computer interaction (ACI): changing perspective on HCI, participation and sustainability. In CHI '13 Extended Abstracts on Human Factors in Computing Systems (CHI EA '13). ACM, New York, NY, USA, 2227-2236. DOI: https://doi.org/10.1145/2468356.2468744

[30] Clara Mancini and Jussi Lehtonen. 2018. The Emerging Nature of Participation in Multispecies Interaction Design. In Proceedings of the 2018 Designing Interactive Systems Conference (DIS '18). ACM, New York, NY, , USA, 907-918. DOI: https://doi.org/10.1145/3196709.3196785

[31] Sean Mealin, Marc Foster, Katherine Walker, Sherrie Yushak, Barbara Sherman, Alper Bozkurt, and David L. Roberts. 2017. Creating an Evaluation System for Future Guide Dogs: A Case Study of Designing for Both Human and Canine Needs. In Proceedings of the Fourth International Conference on Animal-Computer Interaction (ACI2017). ACM, New York, NY, USA, Article 13, 6 pages. DOI: https://doi.org/10.1145/3152130.3152148

[32] Aske Mottelson and Kasper Hornbæk. 2017. Virtual reality studies outside the laboratory. In Proceedings of the 23rd ACM Symposium on Virtual Reality Software and Technology (VRST '17). ACM, New York, NY, USA, Article 9, 10 pages. DOI: https://doi.org/10.1145/3139131.3139141

[33] Ann Morrison, Rune Heide Møller, Cristina Manresa-Yee, and Neda Eshraghi. 2016. The impact of training approaches on experimental setup and design of wearable vibrotactiles for hunting dogs. In Proceedings of the Third International Conference on Animal-Computer Interaction (ACI '16). ACM, New York, NY, USA, Article 4, 10 pages. DOI: https://doi.org/10.1145/2995257.2995391

[34] Steve North. 2017. Hey, where's my hay?: design fictions in horse-computer interaction. In Proceedings of the Fourth International Conference on Animal-Computer Interaction (ACI2017). ACM, New York, NY, USA, Article 17, 5 pages. DOI: https://doi.org/10.1145/3152130.3152149

[35] Steve North, Carol Hall, Amanda Roshier and Clara Mancini. 2015, July. HABIT: Horse automated behaviour identification tool: A position paper. BCS.

[36] Patrizia Paci, Clara Mancini, and Blaine A. Price. 2016. Designing for wearability in animal biotelemetry. In Proceedings of the Third International Conference on Animal-Computer Interaction (ACI '16). ACM, New York, NY, USA, Article 13, 4 pages. DOI: https://doi.org/10.1145/2995257.3012018

[37] Anaïs Racca, Eleonora Amadei, Séverine Ligout, Kun Guo, Kerstin Meints and Daniel Mills. 2010. Discrimination of Human and Dog Faces and Inversion Responses in Domestic Dogs (Canis Familiaris). Anim. Cogn. 2010. 13. 525−533. DOI: https://doi.org/10.1007/s10071-009-0303-3

[38] Janet C. Read and S. MacFarlane. 2006. Using the fun toolkit and other survey methods to gather opinions in child computer interaction. In Proceedings of the 2006 conference on Interaction design and children (IDC '06). ACM, New York, NY, USA, 81-88. DOI: https://doi.org/10.1145/1139073.1139096

[39] Benjamin I. Resner. 2001. Rover@Home: Computer Mediated Remote Interaction between Humans and Dogs. Master's Thesis, Massachusetts Institute of Technology, Cambridge, MA, USA, 2001.

[40] Charlotte L. Robinson, Clara Mancini, Janet van der Linden, Claire Guest, and Robert Harris. 2014. Canine-centered interface design: supporting the work of diabetes alert dogs. In Proceedings of the SIGCHI Conference on Human Factors in Computing Systems (CHI '14). ACM, New York, NY, USA, 3757-3766. DOI: https://doi.org/10.1145/2556288.2557396

[41] Alexandre Pongrácz Rossi, Sarah Rodriguez, and Cassia Rabelo Cardoso dos Santos. 2016. A dog using skype. In Proceedings of the Third International Conference on Animal-Computer Interaction (ACI '16). ACM, New York, NY, USA, Article 10, 4 pages. DOI: https://doi.org/10.1145/2995257.3012019

[42] Patrick C. Shih and Christena Nippert-Eng. 2016. From Quantified Self to Quantified Other: Engaging the Public on Promoting Animal Well-being ACM Classification Keywords. ACM Conference on Human Factors in Computing Systems: Workshop on HCI Goes to the Zoo.

[43] Scott Smith. 2016. Insights: Scott Smith. Medium. Design Friction.

[44] Sanni Somppi, Heini Törnqvist, Laura Hänninen, Christina Krause, and Outi Vainio. 2012. Dogs Do Look at Images: Eye Tracking in Canine Cognition Research. Anim. Cognition. 15. 163−174. DOI: https://doi.org/10.1007/s10071-011-0442-1

[45] Roger Thomas Kok Cheun Tan, Adrian David Cheok, Roshan Lalintha Peiris, I.J.O. Wijesena, Derek Bing Siang Tan, Karthik Raveendra, Khanh Dung Thi Nguyen, Yin Ping Sen, and Elvin Zhiwen Yio. 2007. Computer game for small pets and humans. In Proceedings of the 6th international conference on Entertainment Computing (ICEC'07), Lizhuang Ma, Matthias Rauterberg, and Ryohei Nakatsu (Eds.). Springer-Verlag, Berlin, Heidelberg, 28-38.

[46] Katriina Tiira, Osomo Hakosalo, L.auri Kareinen, Anne Thomas, Anna Hielm-Björkman, Catherine Escriou, Paul Arnold, and Hannes Lohi. 2012. Environmental effects on compulsive tail chasing in dogs. PloS one. 7, 7. DOI: https://doi.org/10.1371/journal.pone.0041684

[47] Giancarlo Valentin, Joelle Alcaidinho, Ayanna Howard, Melody M. Jackson, and Thad Starner. 2015. Towards a canine-human communication system based on head gestures. In Proceedings of the 12th International Conference on Advances in Computer Entertainment Technology (ACE '15). ACM, New York, NY, USA, Article 65, 9 pages. DOI: https://doi.org/10.1145/2832932.2837016

[48] Sarah Webber, Marcus Carter, Sally Sherwen, Wally Smith, Zaher Joukhadar, and Frank Vetere. 2017. Kinecting with Orangutans: Zoo Visitors' Empathetic Responses to Animals? Use of Interactive Technology. In Proceedings of the 2017 CHI Conference on Human Factors in Computing Systems (CHI '17). ACM, New York, NY, USA, 6075-6088. DOI: https://doi.org/10.1145/3025453.3025729

[49] Michelle Westerlaken and Stefano Gualeni. 2016. Becoming with: towards the inclusion of animals as participants in design processes. In Proceedings of the Third International Conference on Animal-

Computer Interaction (ACI '16). ACM, New York, NY, USA, Article 1, 10 pages. DOI: https://doi.org/10.1145/2995257.2995392
[50] Hanna Wirman. 2014. Games for/with strangers-Captive orangutan (pongo pygmaeus) touch screen play. Antennae. DOI: http://hdl.handle.net/10397/7611
[51] Shelten Gee Jao Yuen, James Park, and Atiyeh Ghoreyshi. 2017. User identification via motion and heartbeat waveform data. U.S. Patent 9,693,711.

[52] Clint Zeagler, Jay Zuerndorfer, Andrea Lau, Larry Freil, Scott Gilliland, Thad Starner, and Melody Moore Jackson. 2016. Canine computer interaction: towards designing a touchscreen interface for working dogs. In Proceedings of the Third International Conference on Animal-Computer Interaction (ACI '16). ACM, New York, NY, USA, Article 2, 5 pages. DOI: https://doi.org/10.1145/2995257.2995384